%% file: VLC_NOMA_UAV_ConfFormat.tex
\documentclass[conference]{IEEEtran}
\IEEEoverridecommandlockouts
\usepackage{cite}
\usepackage{amsmath,amssymb,amsfonts}
\usepackage{algorithmic}
\usepackage{graphicx}
\usepackage{textcomp}
\usepackage{xcolor}
\def\BibTeX{{\rm B\kern-.05em{\sc i\kern-.025em b}\kern-.08em
    T\kern-.1667em\lower.7ex\hbox{E}\kern-.125emX}}
\begin{document}

\title{Clustering and Power Allocation for \\UAV-assisted NOMA-VLC Systems: \\A Swarm Intelligence Approach\\
\thanks{This work was supported by the National Research Foundation of Korea (NRF) Grant funded by the Korea Government (MSIT) under Grants NRF-2019R1C1C1006143 and NRF-2019R1I1A3A01060518.}
}

\author{\IEEEauthorblockN{Quoc-Viet Pham$ ^{1} $, Nhu-Ngoc Dao$ ^{2} $, Thien Huynh-The$ ^{3} $, Jun Zhao$ ^{4} $, and Won-Joo Hwang$ ^{5} $}
\IEEEauthorblockA{
	$ ^{1} $Research Institute of Computer, Information and Communication, Pusan National University, Busan, 46241 Korea \\
	$ ^{2} $Department of Computer Science and Engineering, Sejong University, Seoul, 05006 Korea\\
	$ ^{3} $ICT Convergence Research Center, Kumoh National Institute of Technology, Gyeongsangbuk-do, 39177 Korea \\
	$ ^{4} $School of Computer Science and Engineering,	Nanyang Technological University, 50 Nanyang Avenue, 639798 Singapore \\
	$ ^{5} $Department of Biomedical Convergence Engineering, Pusan National University, Busan, 46241 Korea \\
	Email: \{vietpq, wjhwang\}@pusan.ac.kr, nndao@ieee.org, thienht@kumoh.ac.kr, junzhao@ntu.edu.sg
}
}
\maketitle

\begin{abstract}
Integrating unmanned aerial vehicles (UAV) to non-orthogonal multiple access (NOMA) visible light communications (VLC) exposes many potentials over VLC and NOMA-VLC systems. 
In this circumstance, user grouping is of importance to reduce the NOMA decoding complexity when the number of users is large; however, this issue has not been considered in the existing study.
In this paper, we aim to maximize the weighted sum-rate of all the users by jointly optimizing UAV placement, user grouping, and power allocation in downlink NOMA-VLC systems.
We first consider an efficient user clustering strategy, then apply a swarm intelligence approach, namely Harris Hawk Optimization (HHO), to solve the joint UAV placement and power allocation problem. Simulation results show outperformance of the proposed algorithm in comparison with four alternatives: OMA, NOMA without pairing, NOMA-VLC with fixed UAV placement, and random user clustering. 
\end{abstract}


\input{Manuscript.tex}



\end{document}

%% file: Manuscript.tex
\section{Introduction}
\label{Sec:Introduction}
Due to the exponential growth of mobile devices and the flexibility requirement of many new applications, it is necessary for 5G and beyond networks (B5G) to support high data rates and massive connectivity. In this context, unmanned aerial vehicles (UAV), and non-orthogonal multiple access (NOMA), and visible light communications (VLC) are considered as key technologies in B5G \cite{Hong2013Performance, Pham2019ASurvey_MEC}. UAV has found many applications for wireless and communications, e.g., aerial base station and computing server, UAV-assisted information dissemination, and UAV-enabled wireless backhaul. The attractive features of UAV are high mobility, flexibility, and provision of line-of-sight (LoS) connections. NOMA, as indicated by the name, is able to support the simultaneous transmission of multiple users using the same resource blocks \cite{Pham2019CoalitionalGames}. In power-domain NOMA, different power values are assigned to different users, whereas successive interference cancellation (SIC) is performed at the receiver side. Besides UAV and NOMA, VLC has been recently utilized in B5G as a key enabler of many applications and services, e.g., indoor tracking and localization, Internet of Things (IoT), and vehicular communications. Main advantages of VLC are security, easy installation, high data rate, low power consumption, and immunity to electromagnetic interference \cite{Marshoud2016NOMA_VLC}. 

Integrating NOMA to UAV and VLC systems has investigated in many existing works. As the very first study on NOMA-VLC, the authors in \cite{Marshoud2016NOMA_VLC} showed a good interplay between NOMA and VLC, for example, high signal-to-noise ratio (SNR), almost stable channels, and high data rate for users. 
To ensure fairness among users, the work \cite{Yang2017FairNOMA} considered optimizing power allocation of the light emitting diode (LED) to maximize the sum logarithmic data rate in a downlink NOMA-VLC system. The studies in \cite{Zhang2017UserGrouping} and \cite{Janjua2020UserPairing} showed that the power allocation schemes with NOMA are superior to those with orthogonal multiple access (OMA) in terms of spectral efficiency. In \cite{Nasir2019UAV_Enabled}, a successive convex approximation (SCA) algorithm was proposed to solve the max-min rate problem via jointly optimizing bandwidth partitioning, power allocation, UAV altitude, and antenna beamwidth.   
The work in \cite{Seo2019UplinkNOMA} proposed a NOMA random access scheme for uplink UAV communications, and showed that the proposed scheme outperforms the original slotted ALOHA. 
A joint UAV location and user association was considered in \cite{Yang2019PowerEfficient} to minimize the total power consumption.  

Except for our recent work in \cite{Pham2020SumRate}, we are not aware of any literature focusing on exploiting potentials of UAVs to further improve NOMA-VLC systems. In \cite{Pham2020SumRate}, we elaborated that the integration of UAV and NOMA into VLC 
is very promising for the following reasons. First, a UAV-assisted NOMA-VLC system can not only provide illumination but also communication services for multiple users simultaneously, thus enabling massive and ubiquitous connectivity for IoT applications in B5G. 
Second, although the energy efficiency of UAVs can be addressed well through energy harvesting and wireless power transfer, it can be further improved when UAVs with VLC capabilities are utilized for communications rather than UAVs with radio frequency (RF) resources. Third, LoS connections and quality of services (QoS) of users in VLC can be guaranteed and improved by UAVs thanks to the features of high mobility and flexibility. Next, several testbed experiments have been carried out to verify the practicability of UAVs in VLCs \cite{Deng2018Twinkle}. Finally, the simulation results showed that the proposed algorithm for UAV-assisted NOMA-VLC performs much better than OMA and fixed-position (i.e., the position of the LED is fixed and not adjustable) schemes. 

Notwithstanding promising results, there is still a challenge that remains unsolved in our previous work \cite{Pham2020SumRate}. As SIC is performed,
each user needs to decode and remove signals of weaker users (i.e., users with worse channel conditions), and in the worst case, the strongest user needs to decode signals of all the other users. Hence, the complexity of employing SIC techniques increases with the number of users, so it is not suitable for scenarios, where a large number of users need to be supported. In this case, user grouping (also known as user clustering and user pairing in the literature) is of vital importance to reduce the computational complexity of NOMA-VLC systems. 
In this work, we solve this issue by studying a joint optimization problem of UAV placement, user grouping, and power allocation, so as to maximize the weighted sum-rate of all the users. In a nutshell, the contributions and features offered by our work can be summarized as follows.
\begin{itemize}
	\item This work is the first attempt to consider joint UAV placement, user grouping, and power allocation for NOMA-VLC systems. In Section~\ref{Sec:Problem_Formulation}, an optimization problem is formulated to maximize the weighted sum-rate of all the users subject to constraints on QoS requirements of users, non-negativity of the transmitted signal, peak optical intensity, and efficient SIC operations. 
	
	\item This work applies an efficient user grouping strategy. In particular, users are divided into two sets: \textit{near-by} and \textit{far} users, which are sorted out in an ascending sequence. Then, a near-by user (nUser) is paired with a far user (fUser) to make a group and to share an orthogonal bandwidth resource (Section~\ref{Sec:ProposedAlgorithm}). 
	
	\item We propose using a swarm intelligence approach, namely Harris Hawk Optimization (HHO), to solve the joint UAV placement and power allocation problem. It has been shown in our previous work, the HHO algorithm can achieve very competitive performance (Section~\ref{Sec:ProposedAlgorithm}).  
	
	\item In Section~\ref{Sec:Simulation}, the proposed algorithm is then compared with the conventional OMA scheme, NOMA without pairing, NOMA-VLC with fixed UAV altitude, and random user grouping. The proposed algorithm outperforms these alternative schemes in terms of weighted sum-rate. Further, simulation results show the need for user clustering and its joint problem with UAV placement and power allocation in UAV-assisted NOMA-VLC systems. 
\end{itemize}


\begin{figure}[t]
	\centering
	\includegraphics[width=0.90\linewidth]{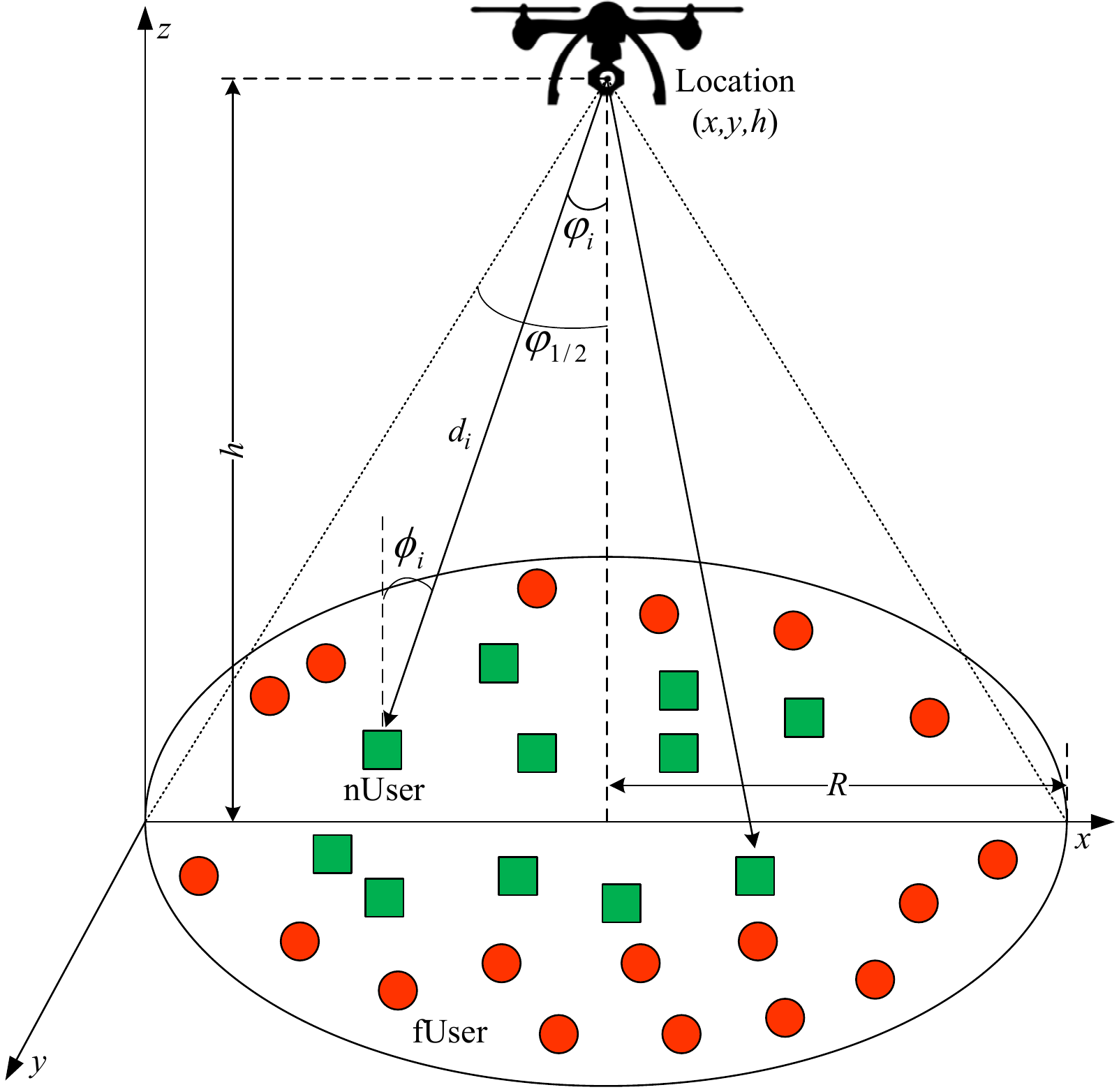}
	\caption{Illustration of a VLC network with UAV and NOMA.}
	\label{Fig:Illustration}	
\end{figure}

\section{Network Setting and Problem Formulation}
\label{Sec:Problem_Formulation}
\subsection{Network Setting}
\label{Subsec:NetworkSetting}
We consider a NOMA-VLC system, as shown in Fig.~\ref{Fig:Illustration}. A UAV is equipped with one LED transmitter (i.e., aerial LAP transmitter/access point (LAP)) to serve $ N $ users, whose positions are randomly created under the coverage area of the aerial LAP transmitter. We note that rotary-wing UAVs have the ability to hover, whereas fixed-wing UAVs need to move forward to remain aloft. In this work, we consider the first type and will study the latter in our future works. 
The set of users is denoted by $ \mathcal{N}=\{1,\dots,N\} $.

This work considers power-domain NOMA, which enables the superposition of signals at the transmitter side, so the aerial LAP can serve multiple users simultaneously using the same time-frequency resource. At the receiver side, users perform SCI to decode and remove messages transmitted for weaker users. We suppose that $ M $ users are selected for making a group, say $ k $-th group, and their channel gains are ordered in an ascending order ($ h_{1k} < \dots < h_{Mk} $), i.e., the first and last users are considered as the weakest user and strongest user, respectively. In case $ M = 2 $, there are two users, including a weak user and a strong user. 
According \cite{Ding2016Impact, Pham2020SumRate}, the achievable rate of the $ i $-th user in $ k $-th group (U$_{ik} $) can be given as
\begin{equation}\label{Eq:DataRate}
R_{ik} = \log_{2}\left(1 + \frac{h_{ik}p_{ik}}{n_{0} + h_{ik}\sum\nolimits_{j = i + 1}^{M}p_{jk}}\right),
\end{equation}
where $ p_{ik} $ is the transmit power for U$_{ik} $ and $ n_{0} $ is the noise power. To ensure SIC operation at the receiver side, the following constraints on power allocation \cite{Ali2016Dynamic}:
\begin{equation}\label{Eq:SIC_Constraint}
p_{ik}\bar{h}_{i+1}^{k} - \sum\nolimits_{j = i+1}^{M}p_{jk}\bar{h}_{i+1}^{k} \geq \theta, \; i = 1, \cdots, M-1,
\end{equation}
where $ \bar{h}_{i}^{k} = h_{ik}/n_{0} $ and $ \theta $ is the minimum power difference, which makes the signal to be decoded and the remaining non-decoded signals distinguishable. In case $ M = 2 $, \eqref{Eq:SIC_Constraint} becomes $ \bar{h}_{2}^{k}\left(p_{1k} - p_{2k}\right) \geq \theta $. In other words, the weaker user is allocated higher power than the strong user, which is needed to ensure high performance of NOMA \cite{Ding2016Impact}.

According to the Lambertian model \cite[Eq.~(10)]{Kahn1997WirelessIC}, the DC channel gain of U$_{ik} $ can be calculated as follows:
\begin{equation}\label{Eq:ChannelGain}
h_{ik} = \frac{A_{ik}}{d_{ik}^{2}} R_{0}(\varphi_{ik}) T_{s}(\phi_{ik}) g(\phi_{ik}) \cos(\phi_{ik})
\end{equation}
for $ 0 \leq \phi_{ik} \leq \Phi_{ik} $ and $ h_{ik} = 0 $ otherwise. In \eqref{Eq:ChannelGain}, $ \Phi_{ik} $ is the field of view (FoV) at U$_{ik} $, $ \varphi_{ik} $ is the angle of irradiance,  $ \phi_{ik} $ is the angle of incidence, $ A_{ik} $ is the detection area, $ d_{ik} $ is the distance to the
aerial LAP, and $ T_{s}(\phi_{ik}) $ is the optical filter gain. Here, $ g(\phi_{ik}) $ is the concentrator gain and is given as
\begin{equation}\label{Eq:concentratorGain}
g(\phi_{ik}) = \frac{q^{2}}{\sin^{2}\Phi_{ik}}, \; 0 \leq \phi_{ik} \leq \Phi_{ik},
\end{equation}
where $ q $ is the refractive index. $ R_{0}(\varphi_{ik}) $ is the radiant intensity and is given as follows:
\begin{equation}\label{key}
R_{0}(\varphi_{ik}) = \frac{\nu+1}{2\pi} \cos^{\nu}\varphi_{ik}
\end{equation}
where $ \nu = -\ln 2 / \ln\left(\cos \varphi_{1/2}\right) $ with $ \varphi_{1/2} $ being the transmitter semiangle at half power. In this paper, we set  $ \varphi_{1/2} = 60^{\circ} $, i.e., $ \nu = 1 $.

As VLC has potentials to provide accurate localization services \cite{Keskin2018Localization}, so in this work we assume that the aerial LAP knows perfectly the locations of GUs, as it has been considered in the literature \cite{Yang2017FairNOMA, Zhang2017UserGrouping}. Denote by $ h $ the fixed altitude of the aerial LAP, and by  $ (x_{u},y_{u},h) $ and $ (x_{ik},y_{ik},0) $ are the coordinates of the aerial LAP and U$_{ik} $, respectively. The distance $ d_{ik} $ can be computed as $ d_{ik} = \sqrt{\left( x_{u} - x_{ik} \right)^{2} + \left( x_{u} - x_{ik} \right)^{2} + h^{2}} $. From Fig.~\ref{Fig:Illustration}, we have $ \cos \varphi_{ik} = \cos \phi_{ik} = h/d_{ik}  $, which can be used to calculate the angles of irradiance and incidence. 

\subsection{Problem Formulation}
\label{Subsec:ProblemFormulation}
To reduce the complexity of users in performing SIC, we assume that two users are paired together to form a cluster and share the same time-frequency resources. Therefore, the number of clusters $ K $ is $ K = N/2 $ and $ M = 2 $. Denote by $ \mathcal{K} $ the set of clusters and by $ \mathcal{M} = \{1,2\} $ the set of users in each cluster. The general case, where each cluster consists of more than two users, is straightforward and easy to be extended from our work. 
Typically, NOMA users have joint benefit to form the grand cluster \cite{Ding2016Impact, Wang2019UserClustering}, i.e., a cluster of all the users. However, as aforementioned, the formation of the grand cluster with a large number of users is not practical owing to the high decoding complexity at strong users. Therefore, to evaluate the contribution of each user in a cluster to the total objective value, each user U$ _{ik} $ is assigned a weight 
as $ \eta_{ik} = K \times \text{O}_{ik}^{-1} $,
where $ \text{O}_{ik} $ is the order of U$ _{ik} $. For example, when $ K = 1 $, $ M = 3 $, $ k = 1 $, we have $ \eta_{11} = 1 $, $ \eta_{21} = 1/2 $, and $ \eta_{31} = 1/3 $. This is reasonable since a strong user needs to perform more SIC operations than a weak user, e.g., U$_{21}$ needs to decode and remove message transmitted for U$_{11}$, whereas U$_{11}$ does not need to perform SIC since it treats messages for U$_{21}$ and U$_{31}$ as noise. 

We consider optimizing the UAV placement, user grouping, and power allocation in order to maximize the total weighted sum-rate. For given user grouping, the optimization problem can be mathematically formulated as follows:
\begin{subequations}
	\label{P1}
	\begin{align}
	& \underset{\{\boldsymbol{w},\boldsymbol{p}\}}{\max}
	& & \sum\limits_{k = 1}^{K}\sum\limits_{i = 1}^{M}\eta_{ik}\log_{2}\left(1 + \frac{h_{ik}p_{ik}}{n_{0} + \sum\nolimits_{j = i+1}^{M}h_{i}p_{j}}\right) \label{P1:a} \\
	& \text{s.t.}
	& & p_{ik} \geq 0, \forall k \in \mathcal{K}, i \in \mathcal{M}, \label{P1:b} \\
	&&& \sum\limits_{k = 1}^{K}\sum\limits_{i = 1}^{M}p_{ik} \leq P_{\max}, \label{P1:c} \\
	&&& \bar{h}_{2}^{k}\left(p_{1k} - p_{2k}\right) \geq \theta, \; \forall k \in \mathcal{K}, \label{P1:e}\\
	&&& R_{ik} \geq R_{ik}^{\text{req}}, \; \forall k \in \mathcal{K}, i \in \mathcal{M}, \label{P1:f} \\
	&&& x_{u}^{2} + y_{u}^{2} \leq R^{2}, \label{P1:g}
	\end{align}
\end{subequations} 
where $ \boldsymbol{w} = \{x_{u},y_{u}\} $, $ \boldsymbol{p} = \{p_{i}\}_{i=1}^{2} $, and $ C = \delta^{-1}\min\{A, B-A\} $, and $ P_{\max} $ denotes the maximal transmit power of the aerial LAP. Constraint~\eqref{P1:e} shows conditions of power allocation for properly performing SIC in NOMA, \eqref{P1:f} imposes QoS requirements of users with $ R_{ik}^{\text{req}} $ being the minimum rate required by U$_{ik} $, and \eqref{P1:g} implies that the UAV should be positioned within a disc with the radius of $ R $. 
In addition, power allocation of the aerial LAP should satisfy the following constraints to maintain the positive intensity of optical signal, and protect eye safety: 
\begin{align} \label{Eq:PositiveIES}
\sum\limits_{k = 1}^{K}\sum\limits_{i = 1}^{M}\sqrt{p_{ik}} \leq \frac{A}{\delta}, \qquad \sum\limits_{k = 1}^{K}\sum\limits_{i = 1}^{M}\sqrt{p_{ik}} \leq \frac{B-A}{\delta}.
\end{align}
Here, $ A $ is a direct current (DC) offset, $ \delta $ is a coefficient, whose value depends on the order of pulse amplitude modulation (PAM), and $ B $ is the peak optical intensity \cite{Kahn1997WirelessIC, Yang2017FairNOMA}. 


Considering the structure of~\eqref{P1}, there are several challenges of finding the solution optimally.
\begin{itemize}
	\item As the sum-rate objective is considered, \eqref{P1} is a non-convex NP-hard in general \cite{Pham2017Fairness}. Moreover, \eqref{P1} is also a mixed-integer non-linear programming (MINLP) problem as both continuous (transmit power and UAV placement) and binary (user grouping) variables are evolved. As a result, it is difficult to obtain the optimal solution in polynomial time.
	
	\item Even with given user grouping, \eqref{P1} is still hard to be solved. The main reason is that the channel gain $ h_{ik} $, as calculated in~\eqref{Eq:ChannelGain}, is difficult to be convexified and/or approximated as convex function with respect to $ d_{ik} $. Therefore, existing approaches like the path-following algorithm in \cite{Nasir2019UAV_Enabled} and success convex approximation (SCA) in \cite{Zhao2019JointTrajectory} are all not applicable.  
	
	\item Existing studies normally adopt a decomposition method to decouple the original problem into transmit power and UAV placement problems, which are then typically solved approximately and iteratively in an alternative manner. In contrast, the HHO algorithm considered in our work can solve the joint problem simultaneously. 
\end{itemize}
In the following section, we will present our proposed algorithm to solve the optimization problem~\eqref{P1}.

\section{Proposed Algorithm}
\label{Sec:ProposedAlgorithm}
As it is difficult to obtain the optimal solution for the problem~\eqref{P1}, we devise an efficient algorithm by decoupling the original problem into two parts: user grouping, and joint placement and power allocation. For user clustering, we propose using an efficient approach, whereas the HHO is utilized to solve the joint UAV placement and power allocation. 

\subsection{User Grouping}
\label{SubSec:UserGrouping}
For VLC-related services, we can reasonably assume that the locations of users are given and do not change during the course of optimization. Intuitively, the aerial LAP should hover at a place, where it can serve most users with reasonable channel qualities. Therefore, in order to perform user paring, the UAV placement is specified as follows:
\begin{equation}
\hat{x}_{u} = \frac{1}{N}\sum\nolimits_{i = 1}^{N}x_{i}, \qquad
\hat{y}_{u} = \frac{1}{N}\sum\nolimits_{i = 1}^{N}y_{i}.
\end{equation}
Then, the UAV coordinate is $ (\hat{x}_{u},\hat{y}_{u},h) $.

Given the coordinates of UAV and users, the channel gains are calculated according to the channel modeling in~\eqref{Eq:ChannelGain}. Based on the sorted set (in an ascending order) of channel gains, all the users are divided into two sets: \textit{nUsers} and \textit{fUsers}. The first-order users in \textit{nUsers} and \textit{fUsers} are paired with each other, thus making the first cluster. Then, the second-order users in these two sets are paired to make the second cluster, and so on. Finally, the $ K $-th cluster is formed by two strongest users.
In doing so, discrepancy in channel gain between any two users can remain apparent, so the performance gain of NOMA over OMA can be achieved \cite{Ding2016Impact}. We note that the main purpose of this work is to show the necessity of user clustering in UAV-assisted NOMA-VLC systems, and finding the optimal user paring strategy will be considered in future. 


\subsection{HHO for Joint UAV Placement and Power Allocation}
\label{SubSec:HHO}

\subsubsection{Overview of HHO}
The HHO is one of the most recent population-based metaheuristics and swarm intelligence techniques, which has become popular and effective in solving many engineering problems since its proposal in \cite{Heidari2019HHO}. The HHO was proposed by Heidari \emph{et al.} in 2019 and mimics the cooperative hunting mechanism of Harris Hawks. 
This algorithm consists of three phases: exploration, exploitation, and transition. For exploitation, the HHO considers four situations, including soft besiege, hard besiege, soft besiege with progressive rapid dives, and hard besiege with progressive rapid dives, which are selected based on the escaping energy and escaping probability of the prey. Due to space limitation, we invite readers to read the seminal paper in \cite{Heidari2019HHO} and our work in \cite{Pham2020SumRate} for more detail of the HHO. In our previous work \cite{Pham2020SumRate}, we showed that HHO can effectively solve the problem of UAV placement and power allocation, and provide very competitive performance.  

\subsubsection{Joint UAV Placement and Power Allocation}
Originally, the HHO is proposed for solving unconstrained optimization problems. To solve the joint UAV placement and power allocation, we propose transforming the constrained problem~\eqref{P1} into an unconstrained one by employing the penalty method. 

Let $ \boldsymbol{X} $ be the solution vector composed of UAV placement and power allocation, and define an indicator function $ H(\cdot) $, which $ H(x) = 1 $ if $ x > 0 $ and $ H(x) = 0 $ if $ x \leq 0 $. 
The penalty value $ P(\boldsymbol{X}) $ can be given as follows: 
\begin{align}\label{Eq:penaltyTerm}
P(\boldsymbol{X}) = 
& -\mu \left(x_{1}\right)^{2} H\left(x_{1}\right) - \mu \left(x_{2}\right)^{2} H\left(x_{2}\right) \notag\\
& - \sum\limits_{k = 1}^{K}\mu \left( x_{3}^{k} \right)^{2}  H\left( x_{3}^{k} \right) - \sum\limits_{i = 1}^{M}\sum\limits_{k = 1}^{K}\mu \left(x_{4}^{ik}\right)^{2} H\left(x_{4}^{ik}\right) \notag\\
& - \mu \left(x_{u}^{2} + y_{u}^{2} - R^{2}\right)^{2} H\left(x_{u}^{2} + y_{u}^{2} - R^{2}\right),
\end{align}
where $ x_{1} = \sum\limits_{i = 1}^{M}\sum\limits_{k = 1}^{K}p_{ik} - P_{\max} $, $ x_{2} = \sum\limits_{i = 1}^{M}\sum\limits_{k = 1}^{K}\sqrt{p_{ik}} - C $, $ x_{3}^{k} =\bar{h}_{2}^{k}\left(\theta - p_{1k} + p_{2k}\right) $ for $ k \in \mathcal{K} $, and $ x_{4}^{ik} = R_{ik}^{\text{req}} - R_{ik} $ for $ i \in \mathcal{M} $ and $ k \in \mathcal{K} $. Here, $ C = \delta^{-1}\min\{A,B-A\} $ and $ \mu $ is a penalty factor. In this paper, we follow the same setting in \cite{Pham2020SumRate}, where $ \mu = 10^{14} $. 
Then, the fitness $ F(\boldsymbol{X}) $ value can be defined as follows:
\begin{equation}\label{Eq:Fitness}
F(\boldsymbol{X}) = \sum\limits_{k = 1}^{K}\sum\limits_{i = 1}^{M}\eta_{ik}\log_{2}\left(1 + \frac{h_{ik}p_{ik}}{n_{0} + \sum\nolimits_{j = i+1}^{M}h_{i}p_{j}}\right) + P(\boldsymbol{X}). 
\end{equation}

\subsection{Computational Complexity}
The computational complexity of the proposed algorithm includes two parts: one from user grouping and the other from the HHO for joint UAV placement and power allocation. The first has a complexity level of $ \mathcal{O}\left(N/2\right) $, whereas the latter is $ \mathcal{O}\left(ST(D + E)\right) $, with $ S, T, D, E $ being the number of hawks, the maximum number of iterations, solution dimensionality, and the number of constraints taken into the penalty term in~\eqref{Eq:penaltyTerm}, respectively. Here, we have $ D = N + 2 $ as there are $ N $ power values corresponding to $ N $ users and $ 2 $ coordinates of the UAV, and $ E = K + MK + 3 $. For simulation in this work, we set $ S = 30 $, $ T = 350 $. From this analysis, the proposed algorithm has the total complexity of $ \mathcal{O}\left(N/2 + ST(2N + 5 + K)\right) $.

\section{Simulation Results}
\label{Sec:Simulation}
This section shows the performance of the proposed algorithm via simulation. We consider a network setting with an aerial LAP. $ N = 20 $ users are randomly distributed in a coverage of 10 m $ \times $ 10 m $ \times $ 3 m. We set $ P_{\max} = 20 $ mW, $ A=20 $, $ B =30 $, and $ n_{0} = -104 $ dBm. Further, we set the bandwidth $ B = 20 $~MHz, $ R = 10 $ m, the detection area of all the users is set to be $ 1 $ cm$ ^{2} $, the optical filter gain is $ 1.0 $, and $ \delta = 3\sqrt{5}/5 $ (for VLC systems with 4-ary PAM). We note that all the plots below are average over 100 realizations, and in each realization, the positions of users are randomly created. 

For comparison with our proposed algorithm (labeled as UPUP), we evaluate four schemes as follows.
\begin{itemize}
	\item \textit{OMA} (labeled as \textbf{OMA}): the bandwidth is divided into $ N $ equal parts and assigned to $ N $ users. In this case, each user is assigned an orthogonal bandwidth resource, so there is no interference among users. 
	
	\item \textit{NOMA without pairing} (labeled as \textbf{cNOMA}): this scheme considers only one cluster, which consists of all the users. This scheme is also the algorithm proposed in our previous work \cite{Pham2020SumRate}. 
	
	\item \textit{Fixed UAV Placement} (labeled as \textbf{fixedP}): this scheme selects a random $ (\boldsymbol{w},h) $, where $ \lVert \mathbf{w} \rVert_{2} =  R^{2}$. This point is then fixed as the coordinate of UAV, and the HHO is used to solve power allocation only. 
	
	\item \textit{Random user grouping} (labeled as \textbf{rClustering}): each cluster is formed by two users at random. After that, we employ the same approach as in ours to solve joint UAV placement and power allocation.
\end{itemize}

\begin{figure}[t]
	\centering
	\includegraphics[width=0.95\linewidth]{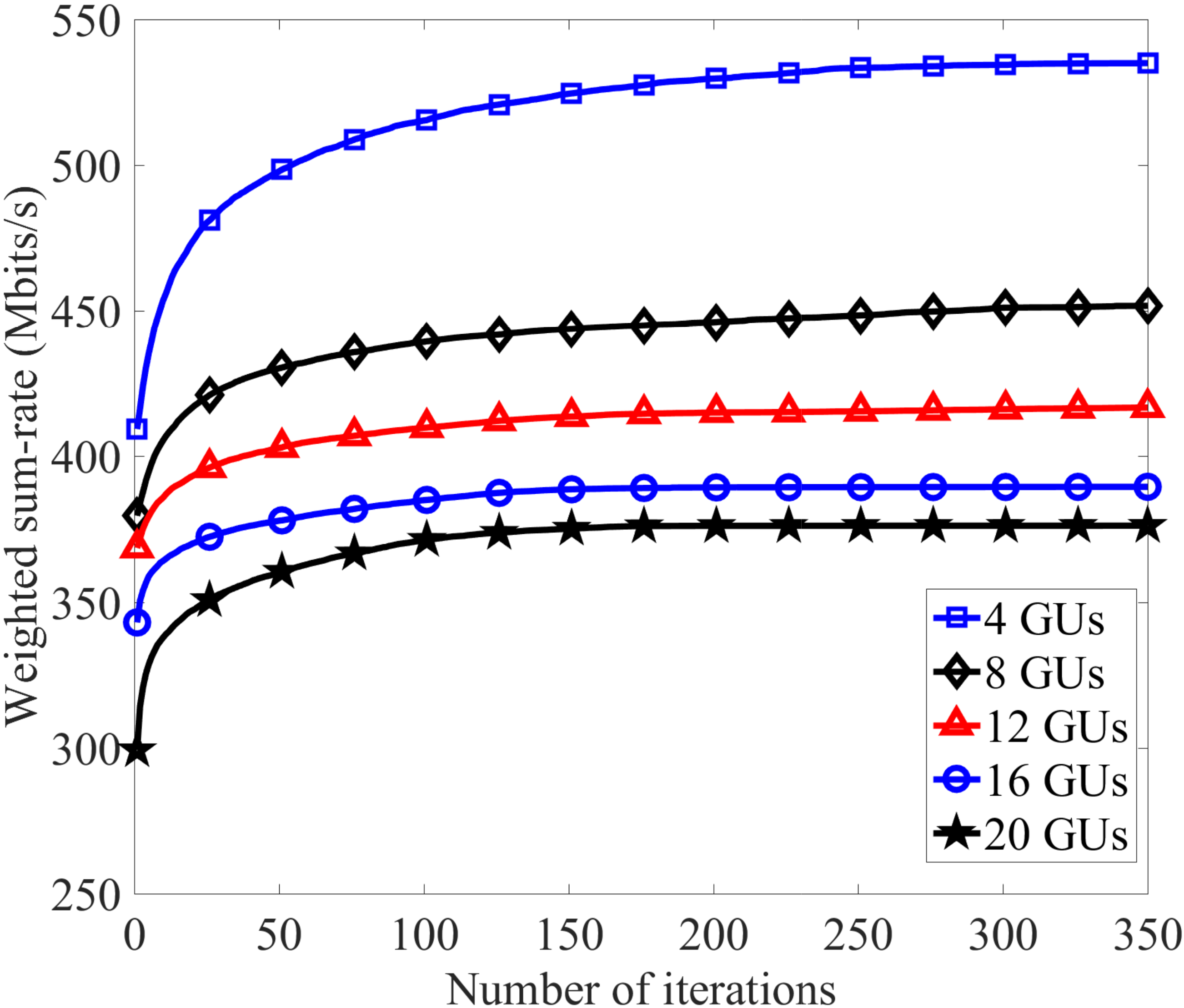}
	\caption{Convergence of the proposed UPUP algorithm.}
	\label{Fig:convergence}	
\end{figure}

First, we evaluate the convergence of the proposed algorithm while varying the number of users. From Fig.~\ref{Fig:convergence}, we can observe that the algorithm converges to the final solution after about 350 iterations for all the cases. The figure also shows the larger the number of users is, the smaller the weighted sum-rate can be achieved. The reason for this is that the amount of bandwidth resources allocated to each cluster gets smaller when the number of users is larger. However, when the number of users is sufficiently large, the bandwidth for each cluster becomes almost the same. In this case, the reduction in the weighted sum-rate becomes smaller. For example, when $ N = 4 $ and $ N = 8 $, each cluster has a bandwidth of $ 10 $ MHz and $ 5 $ MHz respectively. The deviation is therefore $ 5 $ MHz; however, it is only $ 0.5 $ MHz for $ N = 16 $ and $ N = 20 $.
%

%
%
%
%
%
%

\begin{figure}[t]
	\centering
	\includegraphics[width=0.95\linewidth]{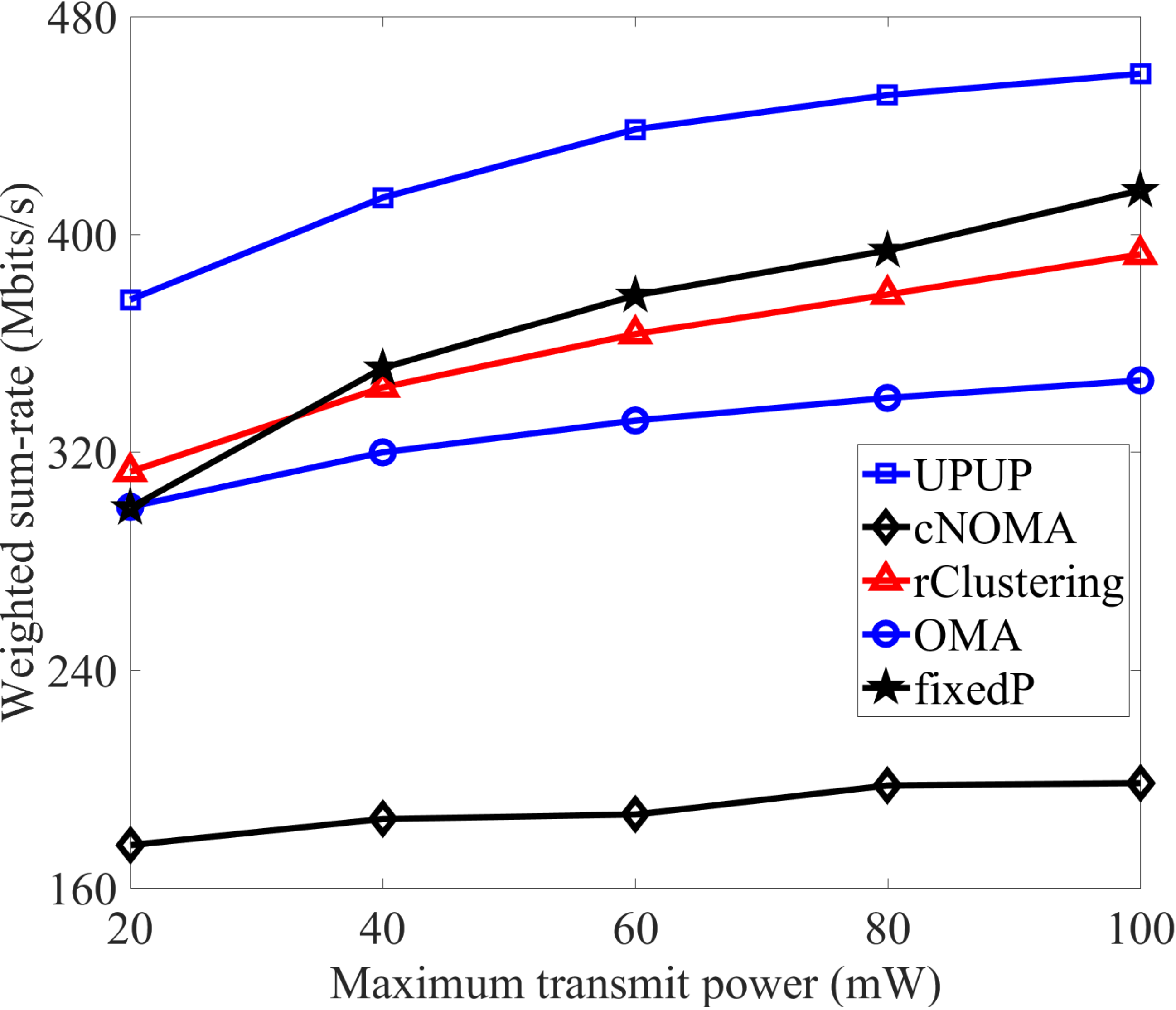}
	\caption{Performance comparison under various $ P_{\max} $.}
	\label{Fig:wsRate_vs_Pmax}	
\end{figure}

Second, we evaluate the performance of the proposed UPUP and alternative algorithm by varying the maximal transmit power $ P_{\max} $ from 20 mW to 100 mW with a deviation step of 20 mW. Fig.~\ref{Fig:wsRate_vs_Pmax} shows that the larger value of $ P_{\max} $ achieves better performance in terms of weighted sum-rate. The reason for this is that larger $ P_{\max} $ enables the aerial LAP to allocate more power to users with good channel conditions when QoS requirements of all the users are already satisfied. An interesting observation is that the OMA scheme is superior to cNOMA for all the $ P_{\max} $ values, which recognizes the importance of user grouping in improving the performance of NOMA-VLC systems. The main reason is from the consideration of decoding complexity and grand coalition in the objective value. It is worth noting that the cNOMA (labeled as HHOPAP in our work \cite{Pham2020SumRate}) still outperforms the OMA scheme when the objective of sum-rate is considered. Another observation is that even with a random clustering strategy, rClustering can yield a better utility than OMA, which shows potentials of NOMA for VLC systems. The figure also shows that our proposed UPUP algorithm outperforms all the other algorithms, including cNOMA, rClustering, OMA, and fixedP. The outperformance of our UPUP algorithm than rClustering justifies the need for jointly optimizing UAV placement rather than joint power allocation and user grouping only. 

\begin{figure}[t]
	\centering
	\includegraphics[width=0.95\linewidth]{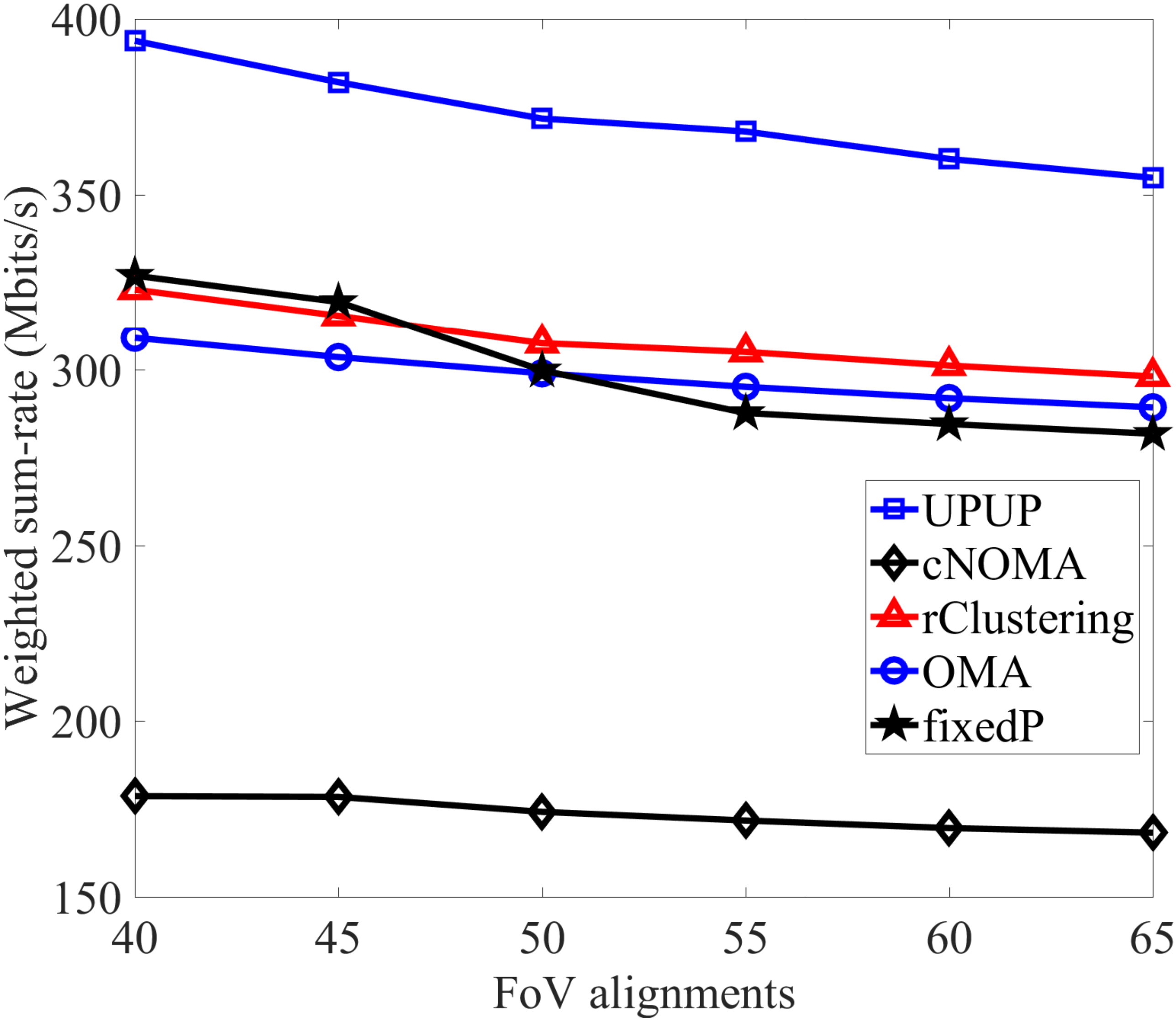}
	\caption{Performance comparison under various fields of view.}
	\label{Fig:wsRate_vs_FoV}	
\end{figure}

\begin{figure}[t]
	\centering
	\includegraphics[width=0.95\linewidth]{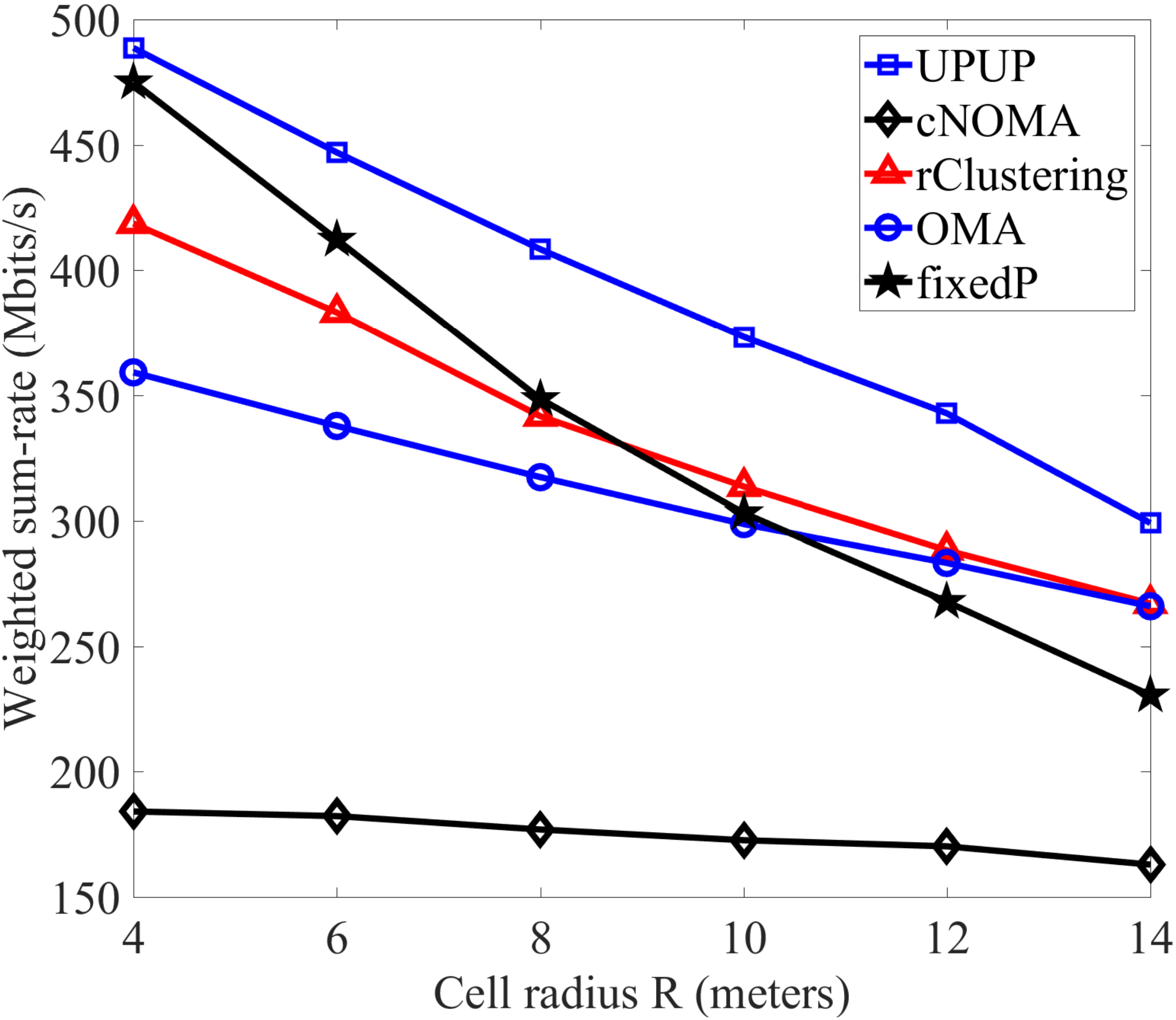}
	\caption{Performance comparison under various cell radii.}
	\label{Fig:wsRate_vs_discR}	
\end{figure}

Next, we vary the FoV of users from 40 to 65 degrees, and compare the performance of the proposed UPUP algorithm with the other schemes. From Fig.~\ref{Fig:wsRate_vs_FoV}, we can observe that the weighted sum-rate increases when the FoV value decreases, i.e., FoV of 45$ ^{o} $ provides higher utility than FoV of 50$ ^{o} $ (under the same simulation settings). As shown in~\eqref{Eq:concentratorGain}, the concentrator gain $ g(\Phi_{ik}) $ gets larger when the $ \Phi_{ik} $ gets smaller. As a note, this work can be further improved if users are equipped with multiple photo-detectors that have different FoV alignments \cite{Hong2013Performance}. It is shown that the proposed UPUP algorithm outperforms all the alternatives. Again, result justifies the need for jointly optimizing user grouping, UAV placement, and power allocation in UAV-assisted NOMA-VLC systems.

Finally, we examine how the weighted sum-rate changes when the cell radius $ R $ varies from 4 to 14 meters. Fig.~\ref{Fig:wsRate_vs_discR} shows that the weighted sum-rate of all the schemes decreases with the cell radius. This is because a large $ R $ shall increase the distance between users and the aerial LAP. Clearly, the proposed algorithm achieves the best performance among all the schemes. The reason is that the proposed approach jointly optimizes UAV placement, user grouping, and power allocation, while the alternatives only optimize a subset of these parameters.

\section{Conclusion} 
\label{Sec:Conclusion}
This paper has considered a joint problem of UAV placement, user grouping, and power allocation in UAV-assisted NOMA-VLC systems. As the problem is difficult to be solved by existing approaches like SCA algorithms, we have used a simple, yet efficient, scheme for user grouping and leveraged the HHO metaheuristic to obtain the solution for joint UAV placement and power allocation. We have conducted simulations to verify that the proposed UPUP algorithm is superior to various existing and baseline algorithms. In the future, we will conduct more simulations under various settings, and consider more sophisticated user grouping schemes to check whether or not the performance can be improved. 

%% file: VLC_NOMA_UAV_ConfFormat.bbl
\begin{thebibliography}{10}
	\providecommand{\url}[1]{#1}
	\csname url@samestyle\endcsname
	\providecommand{\newblock}{\relax}
	\providecommand{\bibinfo}[2]{#2}
	\providecommand{\BIBentrySTDinterwordspacing}{\spaceskip=0pt\relax}
	\providecommand{\BIBentryALTinterwordstretchfactor}{4}
	\providecommand{\BIBentryALTinterwordspacing}{\spaceskip=\fontdimen2\font plus
		\BIBentryALTinterwordstretchfactor\fontdimen3\font minus
		\fontdimen4\font\relax}
	\providecommand{\BIBforeignlanguage}[2]{{%
			\expandafter\ifx\csname l@#1\endcsname\relax
			\typeout{** WARNING: IEEEtran.bst: No hyphenation pattern has been}%
			\typeout{** loaded for the language `#1'. Using the pattern for}%
			\typeout{** the default language instead.}%
			\else
			\language=\csname l@#1\endcsname
			\fi
			#2}}
	\providecommand{\BIBdecl}{\relax}
	\BIBdecl
	
	\bibitem{Hong2013Performance}
	Y.~{Hong}, J.~{Chen}, Z.~{Wang}, and C.~{Yu}, ``Performance of a precoding
	{MIMO} system for decentralized multiuser indoor visible light
	communications,'' \emph{IEEE Photonics Journal}, vol.~5, no.~4, pp.
	7\,800\,211--7\,800\,211, Aug. 2013.
	
	\bibitem{Pham2019ASurvey_MEC}
	Q.-V. Pham, F.~{Fang}, V.~N. {Ha}, M.~J. {Piran}, M.~{Le}, L.~B. {Le}, W.-J.
	Hwang, and Z.~{Ding}, ``A survey of multi-access edge computing in {5G} and
	beyond: Fundamentals, technology integration, and state-of-the-art,''
	\emph{IEEE Access}, vol.~8, pp. 116\,974--117\,017, Jun. 2020.
	
	\bibitem{Pham2019CoalitionalGames}
	Q.-V. Pham, T.~H. {Nguyen}, Z.~{Han}, and W.-J. Hwang, ``Coalitional games for
	computation offloading in {NOMA}-enabled multi-access edge computing,''
	\emph{IEEE Transactions on Vehicular Technology}, vol.~69, no.~2, pp.
	1982--1993, Feb. 2020.
	
	\bibitem{Marshoud2016NOMA_VLC}
	H.~{Marshoud}, V.~M. {Kapinas}, G.~K. {Karagiannidis}, and S.~{Muhaidat},
	``Non-orthogonal multiple access for visible light communications,''
	\emph{IEEE Photonics Technology Letters}, vol.~28, no.~1, pp. 51--54, Jan.
	2016.
	
	\bibitem{Yang2017FairNOMA}
	Z.~{Yang}, W.~{Xu}, and Y.~{Li}, ``Fair non-orthogonal multiple access for
	visible light communication downlinks,'' \emph{IEEE Wireless Communications
		Letters}, vol.~6, no.~1, pp. 66--69, Feb. 2017.
	
	\bibitem{Zhang2017UserGrouping}
	X.~{Zhang}, Q.~{Gao}, C.~{Gong}, and Z.~{Xu}, ``User grouping and power
	allocation for {NOMA} visible light communication multi-cell networks,''
	\emph{IEEE Communications Letters}, vol.~21, no.~4, pp. 777--780, Apr. 2017.
	
	\bibitem{Janjua2020UserPairing}
	M.~B. {Janjua}, D.~B. {da Costa}, and H.~{Arslan}, ``User pairing and power
	allocation strategies for {3D VLC-NOMA} systems,'' \emph{IEEE Wireless
		Communications Letters}, 2020, in press.
	
	\bibitem{Nasir2019UAV_Enabled}
	A.~A. {Nasir}, H.~D. {Tuan}, T.~Q. {Duong}, and H.~V. {Poor}, ``{UAV}-enabled
	communication using {NOMA},'' \emph{IEEE Transactions on Communications},
	vol.~67, no.~7, pp. 5126--5138, Jul. 2019.
	
	\bibitem{Seo2019UplinkNOMA}
	J.~{Seo}, S.~{Pack}, and H.~{Jin}, ``Uplink {NOMA} random access for
	{UAV}-assisted communications,'' \emph{IEEE Transactions on Vehicular
		Technology}, vol.~68, no.~8, pp. 8289--8293, Aug. 2019.
	
	\bibitem{Yang2019PowerEfficient}
	Y.~{Yang}, M.~{Chen}, C.~{Guo}, C.~{Feng}, and W.~{Saad}, ``Power efficient
	visible light communication ({VLC}) with unmanned aerial vehicles ({UAVs}),''
	\emph{IEEE Communications Letters}, vol.~23, no.~7, pp. 1272--1275, Jul.
	2019.
	
	\bibitem{Pham2020SumRate}
	Q.-V. Pham, T.~{Huynh-The}, M.~{Alazab}, J.~{Zhao}, and W.~{Hwang}, ``Sum-rate
	maximization for {UAV}-assisted visible light communications using {NOMA}:
	Swarm intelligence meets machine learning,'' \emph{IEEE Internet of Things
		Journal}, 2020, in press.
	
	\bibitem{Deng2018Twinkle}
	H.~Deng, J.~Li, A.~Sayegh, S.~Birolini, and S.~Andreani, ``Twinkle: A flying
	lighting companion for urban safety,'' in \emph{Proceedings of the Twelfth
		International Conference on Tangible, Embedded, and Embodied Interaction},
	ser. TEI '18.\hskip 1em plus 0.5em minus 0.4em\relax Stockholm, Sweden: ACM,
	2018, pp. 567--573.
	
	\bibitem{Ding2016Impact}
	Z.~{Ding}, P.~{Fan}, and H.~V. {Poor}, ``Impact of user pairing on {5G}
	nonorthogonal multiple-access downlink transmissions,'' \emph{IEEE
		Transactions on Vehicular Technology}, vol.~65, no.~8, pp. 6010--6023, Aug.
	2016.
	
	\bibitem{Ali2016Dynamic}
	M.~S. {Ali}, H.~{Tabassum}, and E.~{Hossain}, ``Dynamic user clustering and
	power allocation for uplink and downlink non-orthogonal multiple access
	({NOMA}) systems,'' \emph{IEEE Access}, vol.~4, pp. 6325--6343, Aug. 2016.
	
	\bibitem{Kahn1997WirelessIC}
	J.~M. {Kahn} and J.~R. {Barry}, ``Wireless infrared communications,''
	\emph{Proceedings of the IEEE}, vol.~85, no.~2, pp. 265--298, Feb. 1997.
	
	\bibitem{Keskin2018Localization}
	M.~F. {Keskin}, A.~D. {Sezer}, and S.~{Gezici}, ``Localization via visible
	light systems,'' \emph{Proceedings of the IEEE}, vol. 106, no.~6, pp.
	1063--1088, Jun. 2018.
	
	\bibitem{Wang2019UserClustering}
	K.~{Wang}, W.~{Liang}, Y.~{Yuan}, Y.~{Liu}, Z.~{Ma}, and Z.~{Ding}, ``User
	clustering and power allocation for hybrid non-orthogonal multiple access
	systems,'' \emph{IEEE Transactions on Vehicular Technology}, vol.~68, no.~12,
	pp. 12\,052--12\,065, Dec. 2019.
	
	\bibitem{Pham2017Fairness}
	Q.-V. Pham and W.-J. Hwang, ``Fairness-aware spectral and energy efficiency in
	spectrum-sharing wireless networks,'' \emph{IEEE Transactions on Vehicular
		Technology}, vol.~66, no.~11, pp. 10\,207--10\,219, Nov. 2017.
	
	\bibitem{Zhao2019JointTrajectory}
	N.~{Zhao}, X.~{Pang}, Z.~{Li}, Y.~{Chen}, F.~{Li}, Z.~{Ding}, and M.~{Alouini},
	``Joint trajectory and precoding optimization for {UAV}-assisted {NOMA}
	networks,'' \emph{IEEE Transactions on Communications}, vol.~67, no.~5, pp.
	3723--3735, May 2019.
	
	\bibitem{Heidari2019HHO}
	A.~A. Heidari, S.~Mirjalili, H.~Faris, I.~Aljarah, M.~Mafarja, and H.~Chen,
	``Harris hawks optimization: Algorithm and applications,'' \emph{Future
		Generation Computer Systems}, vol.~97, pp. 849--872, Aug. 2019.
	
\end{thebibliography}
